\documentclass[preprint,showpacs,preprintnumbers,amsmath,amssymb]{revtex4}
\usepackage{amsmath}

\usepackage{graphicx}% Include figure files
\usepackage{pdfpages}
\usepackage{dcolumn}% Align table columns on decimal point
\usepackage{bm}% bold math
\usepackage{amsmath} 
\usepackage{amsfonts}
\usepackage{amssymb}
\usepackage{url}
\usepackage{microtype}
\usepackage{graphicx}%

\newcommand{\qed}{\nobreak \ifvmode \relax \else
      \ifdim\lastskip<1.5em \hskip-\lastskip
      \hskip1.5em plus0em minus0.5em \fi \nobreak
      \vrule height0.75em width0.5em depth0.25em\fi}

\renewcommand{\vec}[1]{\mathbf{#1}}

\begin{document}

\preprint{}
\title{Quasirandom estimations of  two-qubit operator-monotone-based  separability probabilities}
\author{Paul B. Slater}
 \email{slater@kitp.ucsb.edu}
\affiliation{%
Kavli Institute for Theoretical Physics, University of California, Santa Barbara, CA 93106-4030\\
}
\date{\today}% It is always \today, today,
            
\begin{abstract}
We conduct a pair of quasirandom estimations of the separability probabilities with respect to ten measures on the 15-dimensional convex set of two-qubit states, using its Euler-angle parameterization. The measures include the (non-monotone) Hilbert-Schmidt one, plus nine others based on operator monotone functions. Our results are supportive of  previous assertions that the Hilbert-Schmidt and Bures (minimal monotone) separability probabilities are $\frac{8}{33} \approx 0.242424$ and $\frac{25}{341} \approx 0.0733138$, respectively, as well as  suggestive of the Wigner-Yanase counterpart being $\frac{1}{20}$. However, one result appears   inconsistent (much too small) with an earlier claim of ours that the separability probability associated with the operator monotone (geometric-mean) function $\sqrt{x}$ is 
$1-\frac{256}{27 \pi ^2} \approx 0.0393251$. But a seeming explanation for this disparity is that the volume of states for the $\sqrt{x}$-based measure is infinite. So, the validity of the earlier conjecture--as well as an alternative one, $\frac{1}{9} \left(593-60 \pi ^2\right) \approx 0.0915262$, we now introduce--can not be  examined through the numerical approach 
adopted, at least perhaps not without some truncation procedure for extreme values.
\end{abstract}
 

\maketitle
\section{Introduction}
In our previous paper, ``Master Lovas--Andai and equivalent formulas verifying the $\frac{8}{33}$ two-qubit Hilbert--Schmidt separability probability and companion rational-valued conjectures" \cite[sec. 7.3]{slater2018master}, it was argued that the two-qubit separability probability \cite{ZHSL} based on the measure provided by the operator monotone (geometric-mean) function $f(x)=\sqrt{x}$ would be (with the random-matrix-theoretic Dyson-index $d$ set to 2) given by the ratio 
\begin{equation} \label{sepX1}  
\mathcal{P}_{sep.\sqrt{x}}(\mathbb{C})=\frac{\int\limits_{-1}^1\int\limits_{-1}^x  \tilde{\eta}_{d} \left(
\left.\sqrt{\frac{1-x}{1+x}}\right/ \sqrt{\frac{1-y}{1+y}}	
	\right)\left(1-x^2\right)^{-d/4} \left(1-y^2\right)^{-d/4} (x-y)^d \mbox{d} y\mbox{d} x}{\int\limits_{-1}^1\int\limits_{-1}^x \left(1-x^2\right)^{-d/4} \left(1-y^2\right)^{-d/4} (x-y)^d \mbox{d} y \mbox{d} x}=
\end{equation}
\begin{equation} \label{sqrtxconjecture}
\frac{\frac{\pi ^2}{2}-\frac{128}{27}}{\frac{\pi^2}{2}}= 1-\frac{256}{27 \pi ^2} \approx 0.0393251.
\end{equation}
(A twofold change-of-variables--as in \cite[Thm. 2]{lovasandai}--is employed for the integrations. At the end of this paper, we  introduce an alternative hypothesis ((\ref{sepX2}), (\ref{sqrtxconjecture2})), as well.)  The symmetric and normalized forms of operator monotone functions $f(x)$ satisfy the 
relation $f(x)=x f(\frac{1}{x})$, with the associated measure (volume form) on the $n \times n$ density matrices $D$ being given by $\sqrt{\det(g_f(D))}=\frac{1}{\sqrt{\det(D)}}\Big(2^{\frac{1}{2} (n-1) n} \Pi_{1 \leq i \leq j \leq n} c_f(\mu_i,\mu_j)\Big)^{d/2}$ 
. Here, the $\mu$'s are the $n$ eigenvalues of $D$ and $c_f(x,y)=\frac{1}{y f(x/y)}$ \cite[eq. (26)]{lovasandai}.

Equation (\ref{sepX1}) can be seen to be a modification (with $-\frac{d}{4}$ replacing $d$ as four of the six exponents) of the formula yielding the asserted (non-operator monotone \cite{ozawa}) Hilbert-Schmidt two-qubit separability probability (again with $d=2=2 \alpha$) \cite[eq. (11)]{slater2018master},
\begin{equation} \label{sepX}  
\mathcal{P}_{sep./HS}(\mathbb{C})=\frac{\int\limits_{-1}^1\int\limits_{-1}^x  \tilde{\chi}_{d} \left(
\left.\sqrt{\frac{1-x}{1+x}}\right/ \sqrt{\frac{1-y}{1+y}}	
	\right)\left(1-x^2\right)^d \left(1-y^2\right)^d (x-y)^d \mbox{d} y\mbox{d} x}{\int\limits_{-1}^1\int\limits_{-1}^x  \left(1-x^2\right)^d \left(1-y^2\right)^d (x-y)^d \mbox{d} y \mbox{d} x}=
\end{equation}
\begin{displaymath}
\frac{\frac{2048}{51975}}{\frac{256}{1575}}=\frac{8}{33} \approx 0.242424.
\end{displaymath}

Now, Lemma 7 in \cite{lovasandai} asserts  in the two-rebit ($d=1$) case that $\tilde{\chi}_{1}(\varepsilon) = \tilde{\eta}_{1}(\varepsilon)$ for $\varepsilon \in [0,1]$, $\varepsilon$ being the singular-value ratio 
\cite[sec. II.A.2]{slater2018qubit}. (The tilde symbol indicates normalization at $\varepsilon=1$.) Also,
prior to the above pair of analyses  in \cite{slater2018master}, Lovas and Andai \cite{lovasandai} were able to  {\it formally} establish for this specific $d=1$ case that these two formulas (\ref{sepX1}) and (\ref{sqrtxconjecture}) yielded $\mathcal{P}_{sep.\sqrt{x}}(\mathbb{R}) \approx 0.26223$ and $\mathcal{P}_{sep./HS}(\mathbb{R})=\frac{29}{64}$. For this purpose, they employed 
\begin{equation} \label{BasicFormula}
\tilde{\chi}_1 (\varepsilon ) = 1-\frac{4}{\pi^2}\int\limits_\varepsilon^1 
\left(
s+\frac{1}{s}-
\frac{1}{2}\left(s-\frac{1}{s}\right)^2\log \left(\frac{1+s}{1-s}\right)
\right)\frac{1}{s}
\mbox{d}  s 
\end{equation}
\begin{displaymath}
 = \frac{4}{\pi^2}\int\limits_0^\varepsilon
\left(
s+\frac{1}{s}-
\frac{1}{2}\left(s-\frac{1}{s}\right)^2\log \left(\frac{1+s}{1-s}\right)
\right)\frac{1}{s}
\mbox{d} s.
\end{displaymath}
We noted in \cite{slater2018master} that 
$\tilde{\chi}_1 (\varepsilon )=\tilde{\eta}_1 (\varepsilon )$ has a closed form,
\begin{equation} \label{poly}
\frac{2 \left(\varepsilon ^2 \left(4 \text{Li}_2(\varepsilon )-\text{Li}_2\left(\varepsilon
   ^2\right)\right)+\varepsilon ^4 \left(-\tanh ^{-1}(\varepsilon )\right)+\varepsilon ^3-\varepsilon
   +\tanh ^{-1}(\varepsilon )\right)}{\pi ^2 \varepsilon ^2},    
\end{equation}
where the polylogarithmic function is defined by the infinite sum
	\begin{equation*}
		\text{Li}_s (z) =
		\sum\limits_{k=1}^\infty 
		\frac{z^k}{k^s},
	\end{equation*}
for arbitrary complex $s$ and for all complex arguments $z$ with $|z|<1$. 

Lovas and Andai also formally established for $d=1,2$ the conjecture of Milz and Strunz \cite{milzstrunz} that the separability probability is constant for both of the indicated measures over the Bloch radii of both subsystems. Further, Slater found evidence that this constancy holds more broadly still, in the Hilbert-Schmidt case--in terms of further {\it Casimir invariants} of higher-dimensional systems \cite{slater2016invariance}. In the Appendix here, we  examine whether or not {\it absolute} separability probabilities  might be similarly constant over the Bloch radii of the  subsystems \cite{slater2009eigenvalues}.

The conjecturally  ($d=2$) also equivalent ``separability functions'' employed in equations (\ref{sepX1}) and (\ref{sqrtxconjecture}) are
\begin{equation}
\tilde{\eta_2}(\varepsilon)  = \tilde{\chi_2}(\varepsilon)  =\frac{1}{3} \varepsilon ^2 \left(4-\varepsilon ^2\right).   
\end{equation}
More generally still,  we have \cite[eq. (70)]{slater2018master}
\begin{equation} \label{Ourformula}
\tilde{\chi_d} (\varepsilon )=
\end{equation}
\begin{displaymath}
\frac{\varepsilon ^d \Gamma (d+1)^3 \,
   _3\tilde{F}_2\left(-\frac{d}{2},\frac{d}{2},d;\frac{d}{2}+1,\frac{3
   d}{2}+1;\varepsilon ^2\right)}{\Gamma \left(\frac{d}{2}+1\right)^2},
\end{displaymath}
where the regularized hypergeometric function is denoted.  
(Admittedly, the chain-of-reasoning leading to these functional expressions--except in the two-rebit [$d=1$] case, due to the results of Lovas and Andai--still lacks the  full rigor one would desire.)

For the two-quater[nionic]bit instance, substituting $d=4$ into (\ref{sepX1}) and employing \cite[eq. (59)]{slater2018master}
\begin{equation} \label{quater}
\tilde{\eta_4} (\varepsilon )=\frac{1}{35} \varepsilon^4 (15 \varepsilon^4 -64 \varepsilon^2+84), 
\end{equation}
we reported  
\cite[eq. (88)]{slater2018master} the ratio of 
$\frac{4 \pi ^2}{3}-\frac{5513}{420}$ to $1.478504859 \times 10^{13}$, yielding (the ``infinitesimal'') result
\begin{equation}
\mathcal{P}_{PPT.\sqrt{x}}(\mathbb{Q})  = 2.2510618339 \times 10^{-15}.
\end{equation}
However, it now appears to us that the denominator is fallacious, and simply evaluates to $\infty$.

For still further extensions of these separability functions from Hilbert-Schmidt to more general induced measures, see \cite{slater2018qubit}. By way of example, for the $d=2$ two-qubit setting with the induced measure parameter $k=1$ (where $k=0$ corresponds to Hilbert-Schmidt measure), we have
an extended formula $\tilde{\chi}_{2,1}(\varepsilon)=\frac{1}{4} \varepsilon ^2 \left(3-\varepsilon ^2\right)^2$,  yielding a separability probability of $\frac{61}{143} =\frac{61}{11 \cdot 13} \approx 0.426573$.
\section{Analyses}
We now report a pair of numerical analyses in which we estimate the two-qubit (that is, $d=2$) separability probabilities associated with the Hilbert-Schmidt measure and nine operator monotone functions \cite{andai2006volume,petzsudar}, among them the $\sqrt{x}$ one already noted, as well as the Bures, Kubo-Mori and Wigner-Yanase \cite{gibilisco2003wigner} ones of strong interest. (Andai has a list from which we drew \cite[sec. 4]{andai2006volume}, and the order of which we largely follow.) 

Though the pair of analyses conducted is certainly strongly supportive of our previous assertions that the  two-qubit Hilbert-Schmidt and Bures separability probabilities are $\frac{8}{33}$ \cite{slaterJModPhys} and $\frac{25}{341}$ \cite{slater2019numerical}, respectively, they do  strongly differ (in being much smaller) from the 
$\mathcal{P}_{sep.\sqrt{x}}(\mathbb{C})= 1-\frac{256}{27 \pi ^2} \approx 0.0393251$ claim.
However, upon further reflection, we suspect that this may be an artifact  of the infinite-volume property \cite{andai2006volume} of the $\sqrt{x}$ measure, which needs to be addressed in a more nuanced numerical manner, if at all possible.

It is of interest to compare and contrast the subject matter and methodologies of the present study with that of two of our papers from 2005, ``Silver mean conjectures for 15-d volumes and 14-d hyperareas of the separable two-qubit systems'' \cite{slaterJGP} and ``Qubit-qutrit separability probability ratios'' \cite{slaterPRA}. These studies employed a different (Tezuka-Faure) approach to quasi-Monte Carlo estimation \cite{giray1} than the quasirandom one here, while obtaining volume and {\it  hyperarea} estimates for various operator monotone-based measures. However, in neither study was the geometric-mean-based measure $f(x)= \sqrt{x}$--of central concern here--examined. Also, issues of {\it absolute} separability probabilities were not studied as they had been in our later 2009 paper, ``Eigenvalues, Separability and Absolute Separability of Two-Qubit States'' \cite{JMP2008}, and in the Appendix below.

To conduct the pair of estimations of ten separability probabilities, we employed the SU(4)-based Euler-angle parameterization \cite{tilma2002parametrization} of the 15-dimensional convex set of two-qubit density matrices.  Though in the past, we have, in fact,  extensively employed this  parameterization in separability probability analyses \cite{slater2009eigenvalues,slaterJGP2,slaterA}, we have more recently  \cite{slater_2012,slater2019,slater2019numerical} relied upon the Ginibre-ensemble approach of Osipov, Sommers and {\.Z}yczkowski for generating random states \cite{osipov}. However, their procedure is designed for Hilbert-Schmidt and Bures measures and not apparently for the other operator monotone measures to be investigated here. (Ginibre ensembles can also be employed for the generation of random density matrices with respect to the extension [$k \neq 0$] of Hilbert-Schmidt to induced measures \cite{Induced}.)

In particular, since we wanted to {\it numerically} investigate our conjecture (\ref{sqrtxconjecture}) as to the value of $\mathcal{P}_{sep.\sqrt{x}}(\mathbb{C})$, it seemed appropriate to revert to the use of the Euler-angle parameterization. Let us further note that in the two-qubit setting, rather than 15 (uniformly-distributed) random numbers (needed for 12 Euler angles [$\alpha_i$, $i=1,\ldots,12$] and 3 eigenvalues [$\lambda_i$, $i=1,2,3$, with $\lambda_4=1-\lambda_1-\lambda_2-\lambda_3$]) at each iteration, in the Ginibre-ensemble approach, the considerably larger numbers of 32 and 64 (normally-distributed) ones are required in the Hilbert-Schmidt and Bures  cases, respectively. On the other hand, in the Euler-angle setting, each realization needs to be weighted by the product of the Haar \cite[eq. (34)]{tilma2002parametrization}
\begin{equation}
 \sin \left(2 \alpha _2\right) \sin \left(\alpha _4\right) \sin ^5\left(\alpha _6\right)
   \sin \left(2 \alpha _8\right) \sin ^3\left(\alpha _{10}\right) \cos ^3\left(\alpha
   _4\right) \cos \left(\alpha _6\right) \cos \left(\alpha _{10}\right)   
\end{equation}
and eigenvalue measures,
\begin{equation}
  \frac{\lambda _1^3 \left(\lambda _1-\lambda _2\right){}^2 \lambda _2^2 \left(\lambda
   _1-\lambda _3\right){}^2 \left(\lambda _2-\lambda _3\right){}^2 \lambda _3
   \left(\lambda _1-\lambda _4\right){}^2 \left(\lambda _2-\lambda _4\right){}^2
   \left(\lambda _3-\lambda _4\right){}^2}{\left(\lambda _1 \lambda _2 \lambda _3
   \lambda _4\right){}^{7/2} f\left(\frac{\lambda _1}{\lambda _2}\right)
   f\left(\frac{\lambda _1}{\lambda _3}\right) f\left(\frac{\lambda _2}{\lambda
   _3}\right) f\left(\frac{\lambda _1}{\lambda _4}\right) f\left(\frac{\lambda
   _2}{\lambda _4}\right) f\left(\frac{\lambda _3}{\lambda _4}\right)},  
\end{equation} 
while in the Ginibre-ensemble alternative, each $4 \times 4$ density matrix produced simply receives equal weight. It would clearly be of interest to evaluate the relative merits of the two methodologies in their common domains of application. 

Further, we used the quasirandom (generalized golden-ratio) estimation methodology recently developed by Martin Roberts \cite{Roberts,Roberts32D,slater2019numerical} with its single free $\alpha_0 \in [0,1]$ parameter set to $\frac{1}{4}$ in one analysis and $\frac{3}{4}$ in the  companion one. At each iteration of these two procedures, we obtain 15 numbers in [0,1]. Interestingly, we were able to jointly use (multiplying by $\pi$ or $\frac{\pi}{2}$, as appropriate) 12 of them for the Euler-angle parameters, and  the other 3 (by sorting them, appending 0 and 1, and taking differences) to obtain the four eigenvalues constrained to sum to 1.
(To greatly speed our computations, we employed the Compile[, CompilationTarget $\longrightarrow$ "C",
 RuntimeAttributes $\longrightarrow$ {Listable}, Parallelization $\longrightarrow$ True] feature of Mathematica, but doing so restricted us to the use of single/normal precision. As the estimation proceeds, and greater successive  are employed as seeds, the occurrence of overflows in the computations noticeably increases. These limited instances have to be discarded, but presumably no systematic effects are introduced by doing so.)
\subsection{Quasirandom procedure}
As noted, we have employed an ``open-ended'' sequence (based on extensions of the  golden ratio \cite{livio2008golden}) recently introduced by Martin Roberts in the detailed presentation ``The Unreasonable Effectiveness 
of Quasirandom Sequences'' \cite{Roberts}.
 
Roberts notes: ``The solution to the 
$d$-dimensional problem, depends on a special constant $\phi_d$, where $\phi_d$ is the value of the smallest, positive real-value of x such that''
\begin{equation}
  x^{d+1}=x+1,
\end{equation}
($d=1$, yielding the golden ratio, and $d=2$, the ``plastic constant'' \cite{Roberts32D}). 
The  $n$-th terms in the quasirandom (Korobov) sequence take the form
\begin{equation} \label{QR}
(\alpha _0+n \vec{\alpha}) \bmod 1, n = 1, 2, 3, \ldots  
\end{equation}
where we have the $d$-dimensional vector,
\begin{equation} \label{quasirandompoints}
\vec{\alpha} =(\frac{1}{\phi_d},\frac{1}{\phi_d^2},\frac{1}{\phi_d^3},\ldots,\frac{1}{\phi_d^d}).  "
\end{equation}
The additive constant $\alpha_0$ is typically taken to be 0. ``However, there are some arguments, relating to symmetry, that suggest that $\alpha_0=\frac{1}{2}$
is a better choice,''  Roberts observes.

In \cite{slater2019numerical}, such points  {\it uniformly} distributed in the $d$-dimensional hypercube $[0,1]^d$, were converted, using an algoirthm of Henrik Schumacher  \cite{Schumacher} to (quasirandomly distributed) normal variates, required for the generation of Ginibre ensembles. However, here, since we rely upon the Euler-angle , such a conversion is not required.
\subsection{Results}
\begin{figure}
    \centering
    \includegraphics{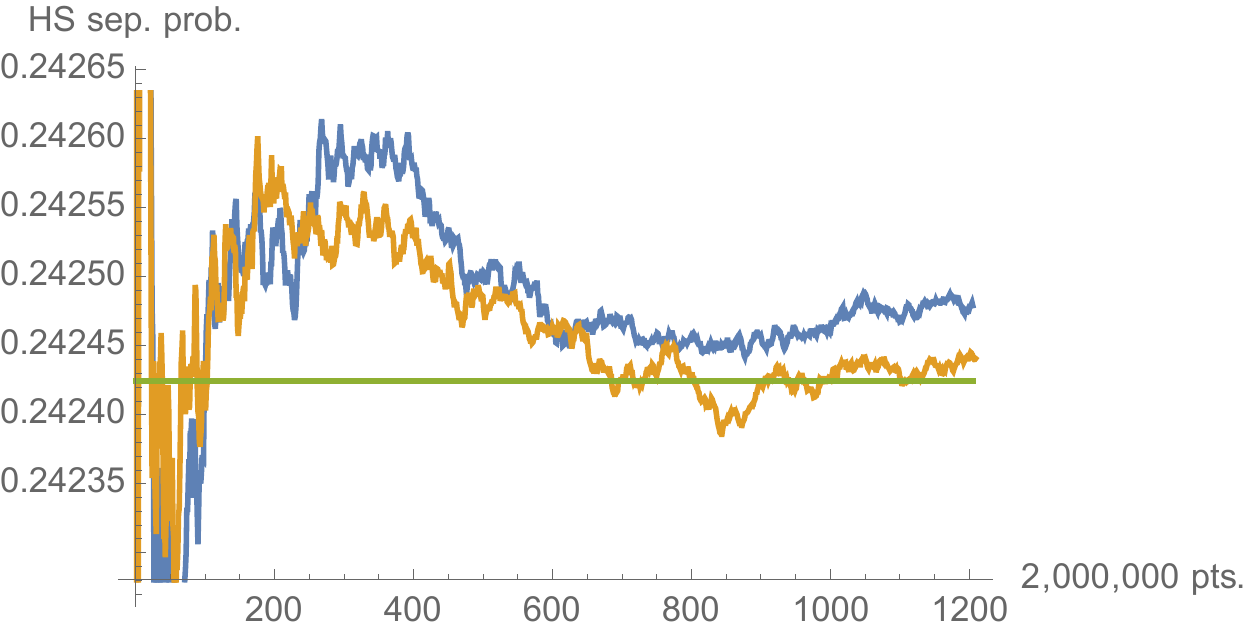}
    \caption{Pair of estimates with respect to the Hilbert-Schmidt measure along with the conjectured value of $\frac{8}{33}$. The $x$ axis here--and in all our figures but the last two are labeled in units of two million points, so the label 1200 corresponds to two billion four hundred million points generated. }
    \label{fig:HS}
\end{figure}

In Fig.~\ref{fig:HS} we show the pair of quasirandom estimates obtained with respect to the Hilbert-Schmidt measure along with the conjectured value of $\frac{8}{33}$  \cite{slaterJModPhys}.
The $x$ axis here--and in all our figures but the last two are labeled in units of two million points, so the label 1200 corresponds to two billion four hundred million points generated. We conducted paired analyses, since it was computationally convenient given the two Mathematica kernels available to us.
The blue (largely greater-valued) curve is based on the Roberts parameter $\alpha_0 =\frac{1}{4}$, and the other (orange) based on $\alpha_0=\frac{3}{4}$. The (arithmetic) average of the last two values is 0.24246.

\begin{figure}
    \centering
    \includegraphics{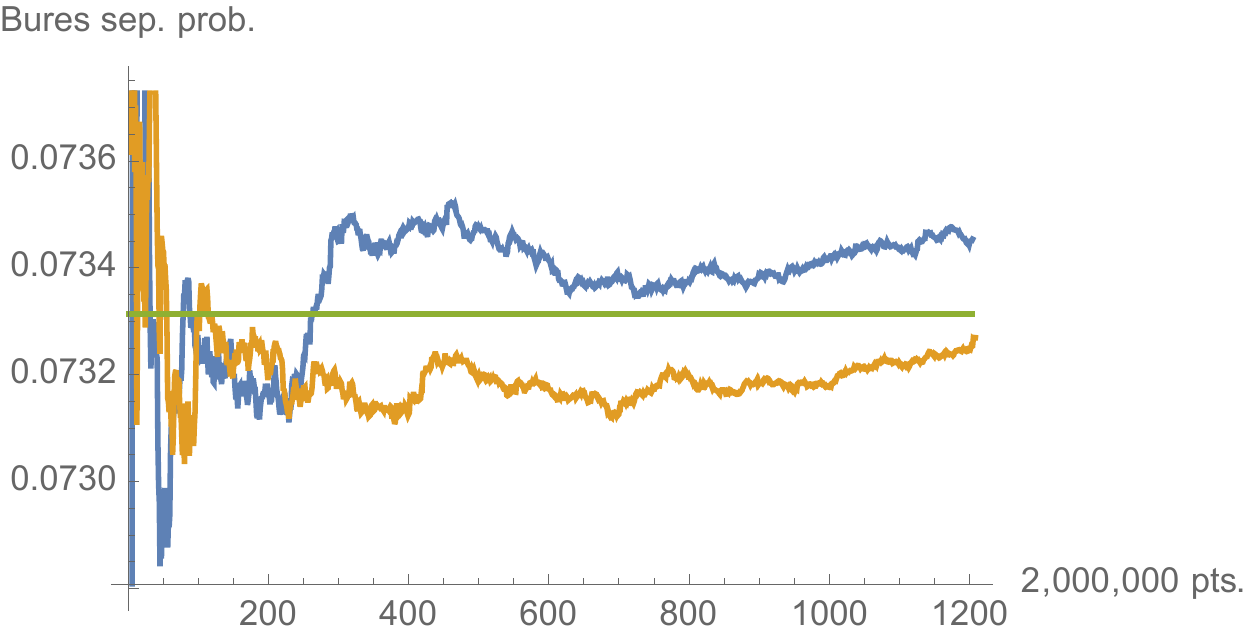}
    \caption{Pair of estimates with respect to the Bures ($f(x)=\frac{x+1}{2}$) measure along with the conjectured value of $\frac{25}{341}$}
    \label{fig:Bures}
\end{figure}

In Fig.~\ref{fig:Bures} we show the pair of estimates with respect to the Bures (minimal monotone) ($f(x)=\frac{x+1}{2}$) measure accompanied by the conjectured value of $\frac{25}{341}$  \cite{slater2019numerical}. 

\begin{figure}
    \centering
    \includegraphics{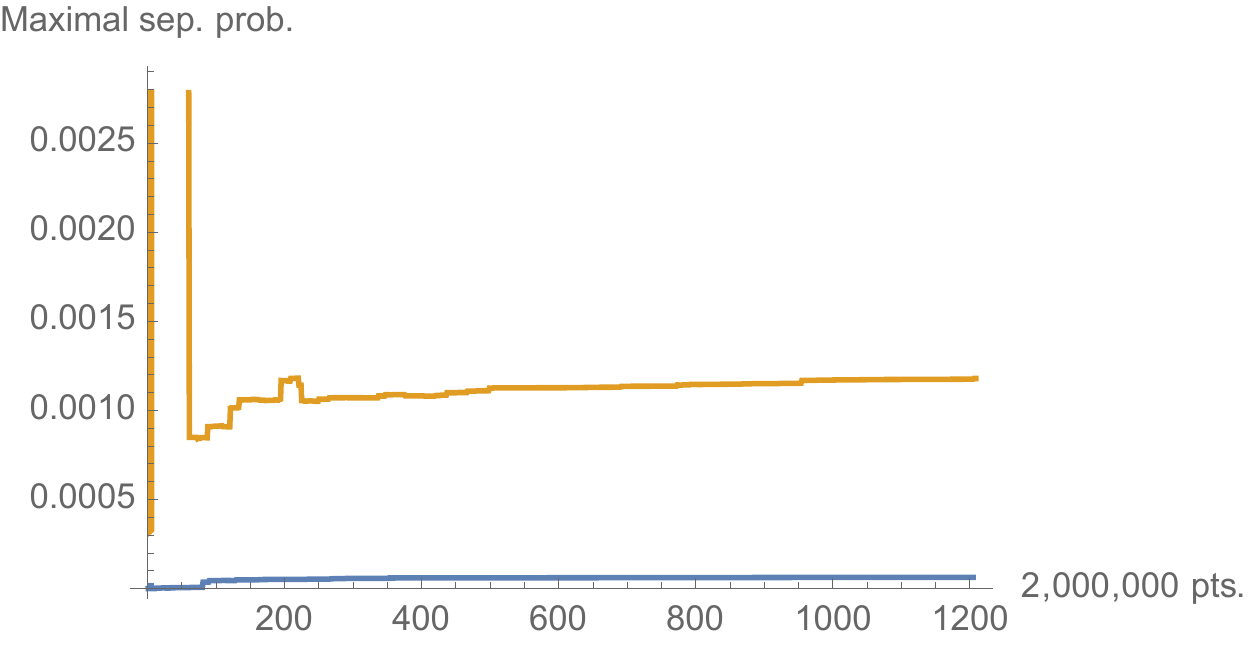}
    \caption{Pair of estimates with respect to the maximal ($f(x)=\frac{2 x}{x+1}$) measure}
    \label{fig:maximal}
\end{figure}

Further, in Fig.~\ref{fig:maximal} we show the pair of (near-zero) estimates with respect to the maximal ($f(x)=\frac{2 x}{x+1}$) measure. The volume of two-qubit states associated with this measure is, however,  apparently infinite \cite[sec. 4]{lovasandai}.

\begin{figure}
    \centering
    \includegraphics{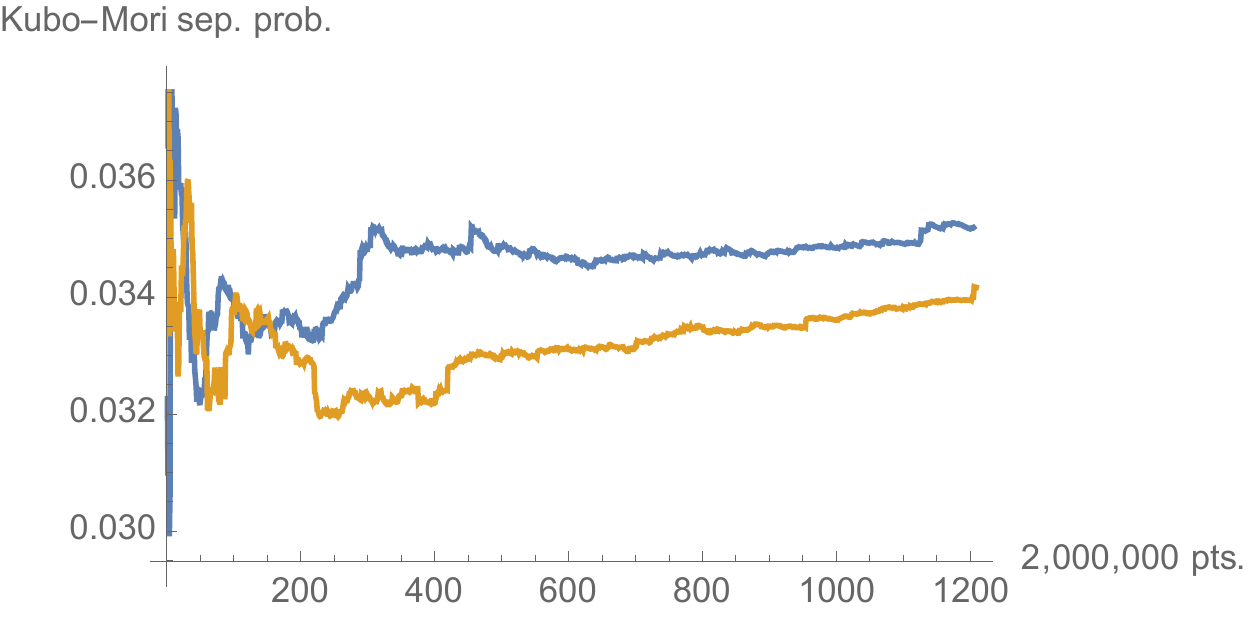}
    \caption{Pair of estimates with respect to the  Kubo-Mori ($f(x)=\frac{x-1}{\log (x)}$) measure}
    \label{fig:Kubo-Mori}
\end{figure}

\begin{figure}
    \centering
    \includegraphics{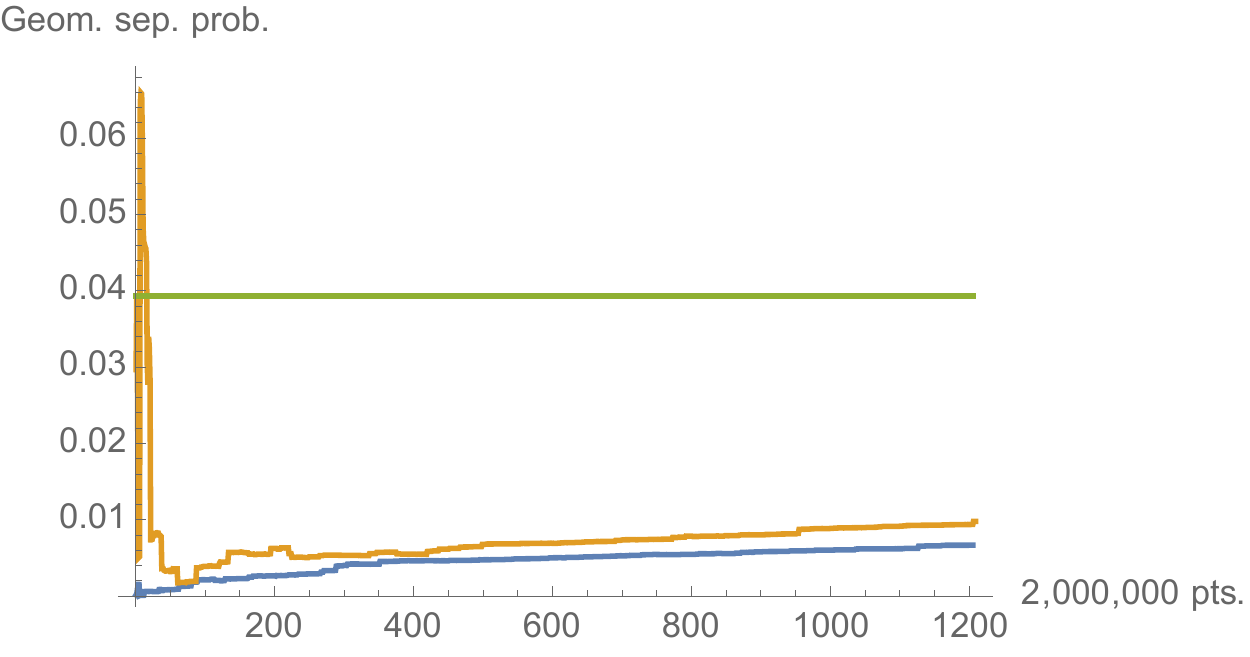}
    \caption{Pair of estimates with respect to the geometric mean ($\sqrt{x}$-based) measure along with the conjectured value of $1-\frac{256}{27 \pi ^2} \approx 0.0393251$--given in (\ref{sqrtxconjecture})}
    \label{fig:sqrt}
\end{figure}

In Fig.~\ref{fig:Kubo-Mori} we show the pair of estimates with respect to the Kubo-Mori ($f(x)=\frac{x-1}{\log (x)}$) measure,
while, in Fig.~\ref{fig:sqrt} we show the pair of estimates obtained using the geometric mean  ($f(x)=\sqrt{x}$) measure. 

This last plot would appear to constitute evidence against the validity of the conjecture that  $\mathcal{P}_{sep.\sqrt{x}}(\mathbb{C}) = 1-\frac{256}{27 \pi ^2} \approx 0.0393251$ given in eq. (\ref{sqrtxconjecture}). However, we must note that a seeming explanation for this inconsistency is that the volume of states for the $\sqrt{x}$-based measure is infinite, as observed by Lovas and Andai 
\cite[sec. 5]{lovasandai}.  Perhaps, a numerical analysis in which a threshold on the magnitude of the $\sqrt{x}$ measure sampled is imposed would be appropriate. Another  strategy might be to require that no randomly generated eigenvalue employed be less than a certain magnitude.  Further, the quite small estimated separability probability  ($\approx 0.005$) in Fig.~\ref{fig:sqrt} is  rather surprising, since 
in the two-rebit ($d=1$) scenario $\mathcal{P}_{sep.\sqrt{x}}(\mathbb{R}) \approx 0.26223$ and  $\mathcal{P}_{sep/HS}(\mathbb{R}) =\frac{29}{64} \approx 0.453125$ are of similar magnitudes.

Relatedly, Lovas and Andai stated--with regard to the $\sqrt{x}$-measure--that ``We show that the volumes of rebit-rebit and qubit-qubit states are infinite, although there is a simple and reasonable method to define the separability
probabilities.  We present integral formulas for separability probabilities in this setting, too.'' Also, they wrote:
``Contrary to the $2\times 2$ case $\ldots$
  the volume of the statistical manifold $(\mathcal{D}_{4,K},g_{\sqrt{x}})$
  is infinite in both of the real and complex cases because $\eta_d (1)=\infty$
   and the
  volume admits the following factorization
\begin{equation*}
\mbox{Vol}_{\sqrt{x}}\mathcal{(D}_{4,\mathbb{K}}).= 4\eta_d (1) \times
  \int_{\mathcal{D}_{2,\mathbb{K}}}\det(D)^{\frac{5}{2}d-\frac{d^2}{2}-1} \mbox{d}\lambda_{d+1}(D) \times \int_{\varepsilon_{2,\mathbb{K}}} \mbox{det} (I-Y^2)^\frac{d-2}{4}\mbox{d} \lambda_{d+2}(Y)".
\end{equation*}
(For further reference, with regard to the alternative hypothesis given in ((\ref{sepX2}), (\ref{sqrtxconjecture2})) below, note the presence of the exponents $\frac{5}{2}d-\frac{d^2}{2}-1$ and $\frac{d-2}{4}$, equalling 2 and 0, respectively, for $d=2$.)

\begin{figure}
    \centering
    \includegraphics{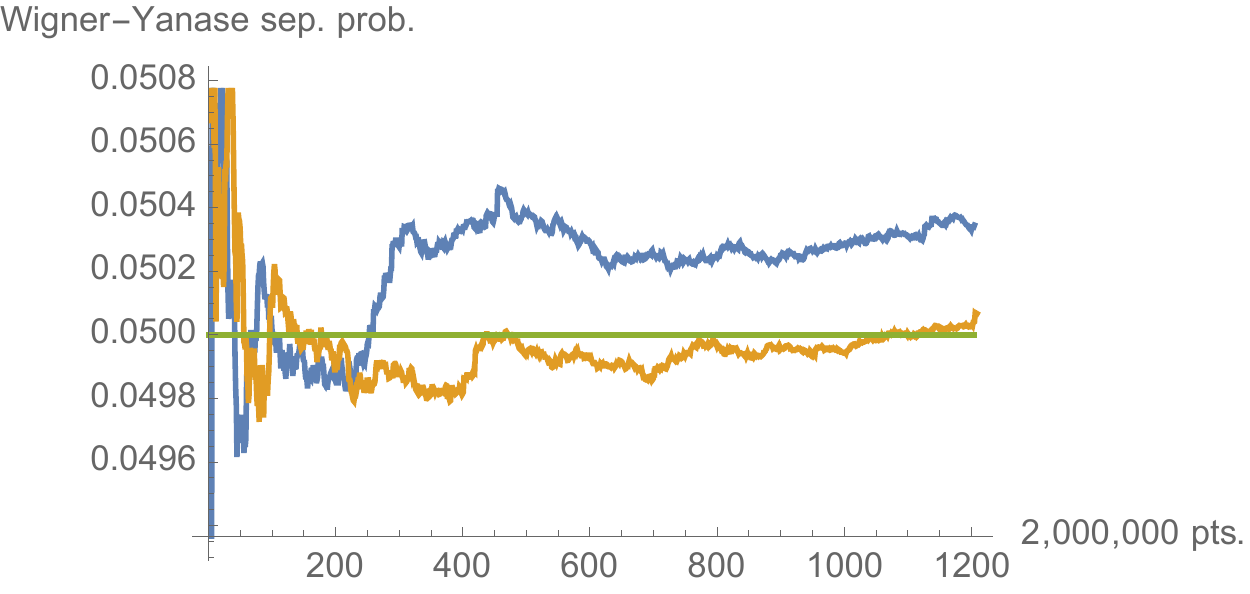}
    \caption{Pair of estimates with respect to the Wigner-Yanase $f(x)=(\frac{1}{4} \left(\sqrt{x}+1\right)^2)$ measure, along with well-fitting $\frac{1}{20}$ line}
    \label{fig:WignerYanase}
\end{figure}

In Fig.~\ref{fig:WignerYanase} we show the pair of estimates (interestingly close to $\frac{1}{20}$) with respect to the Wigner-Yanase ($f(x)=\frac{1}{4} \left(\sqrt{x}+1\right)^2$) measure.  (A third estimation--now with Roberts parameter $\alpha_0=0$  and 316 million 
realizations--also gave us a close estimate  of 0.0499207. Additionally, a fourth [Tezuka-Faure sequence quasi-Monte Carlo] estimate of 0.0503391 was reported in Table II of our 2005 study \cite{slaterJGP}. In that table, estimates of 0.0346801 and 0.0609965 were reported for the Kubo-Mori and identric measures.)

\begin{figure}
    \centering
    \includegraphics{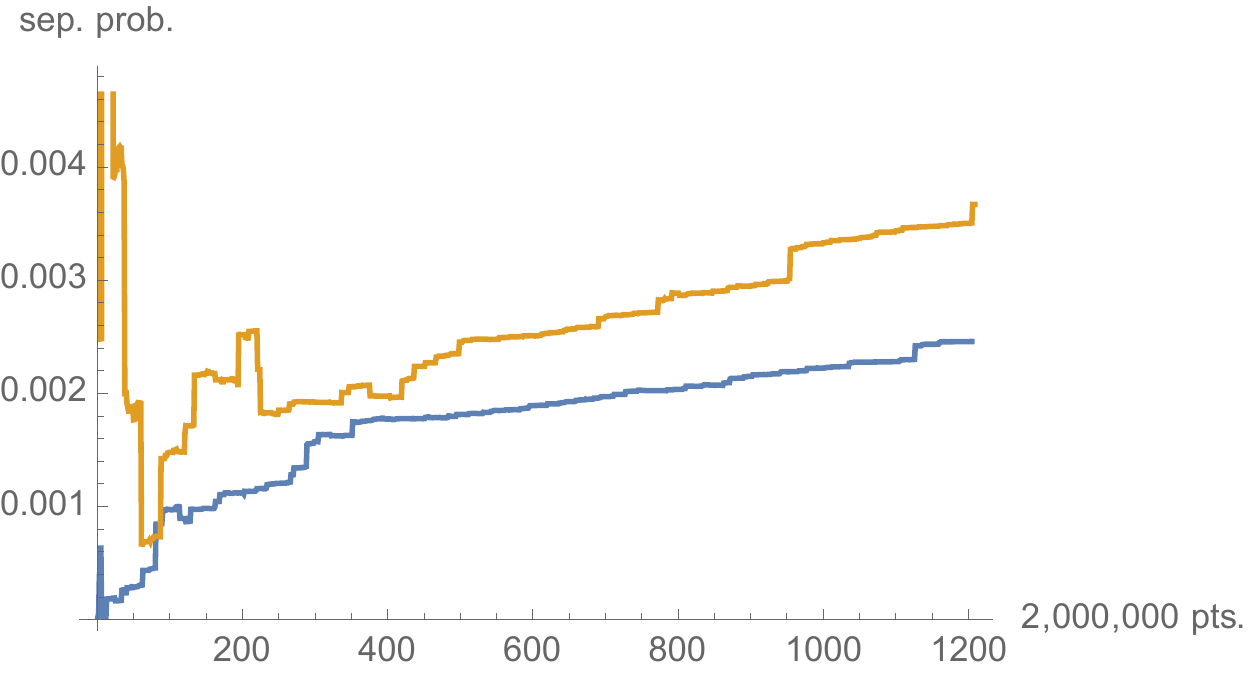}
    \caption{Pair of estimates with respect to the  $f(x)=\frac{2 (x-1) \sqrt{x}}{(x+1) \log (x)}$  measure}
    \label{fig:other1}
\end{figure}

In Fig.~\ref{fig:other1} we present the pair of estimates with respect to the  $f(x)=\frac{2 (x-1) \sqrt{x}}{(x+1) \log (x)}$  measure. Again, the volume of two-qubit states associated with this measure is apparently infinite \cite[sec. 4]{lovasandai}.

\begin{figure}
    \centering
    \includegraphics{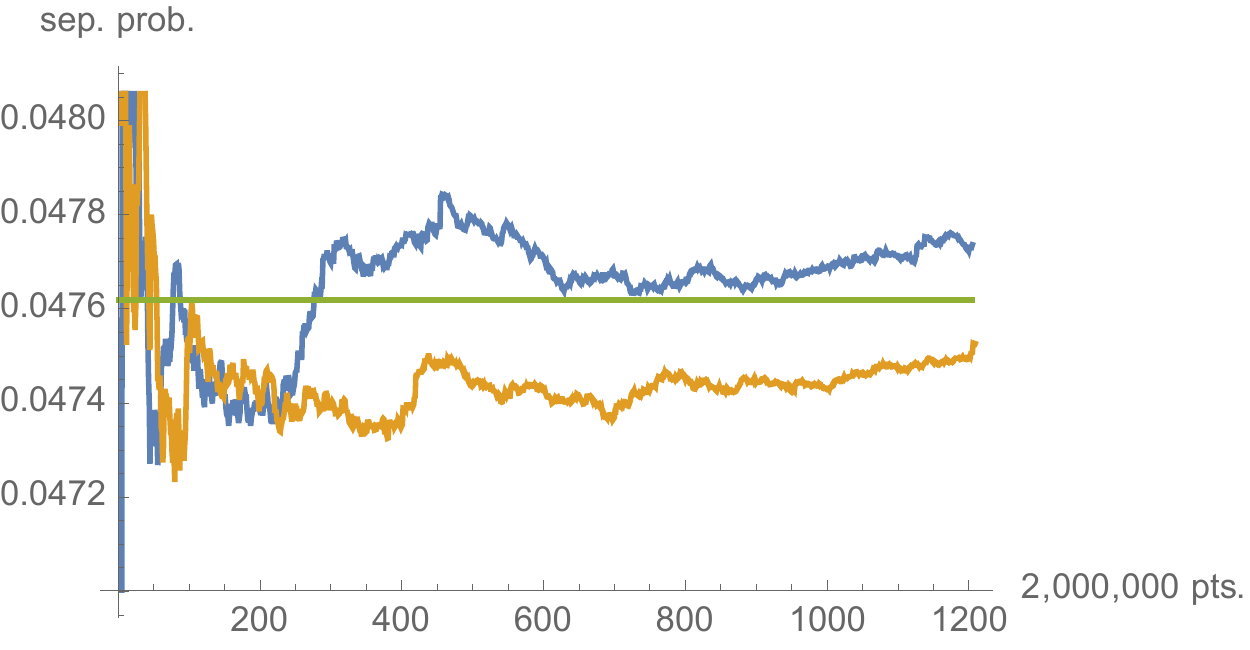}
    \caption{Pair of estimates with respect to the  $f(x)=\frac{x^2+6 x+1}{4 x+4}$ measure,  along with the closely-fitted value of $\frac{1}{21}$}
    \label{fig:ARITH}
\end{figure}

In Fig.~\ref{fig:ARITH} we show the pair of estimates with respect to the  $f(x)=\frac{x^2+6 x+1}{4 x+4}$ measure, along with the closely-fitted value of $\frac{1}{21}$. (This function is the arithmetic average of the ones for the minimal (Bures)--$\frac{x+1}{2}$--and maximal--$\frac{2 x}{x+1}$--measures, as noted in  \cite[eq. (14)]{slaterJGP}.)

\begin{figure}
    \centering
    \includegraphics{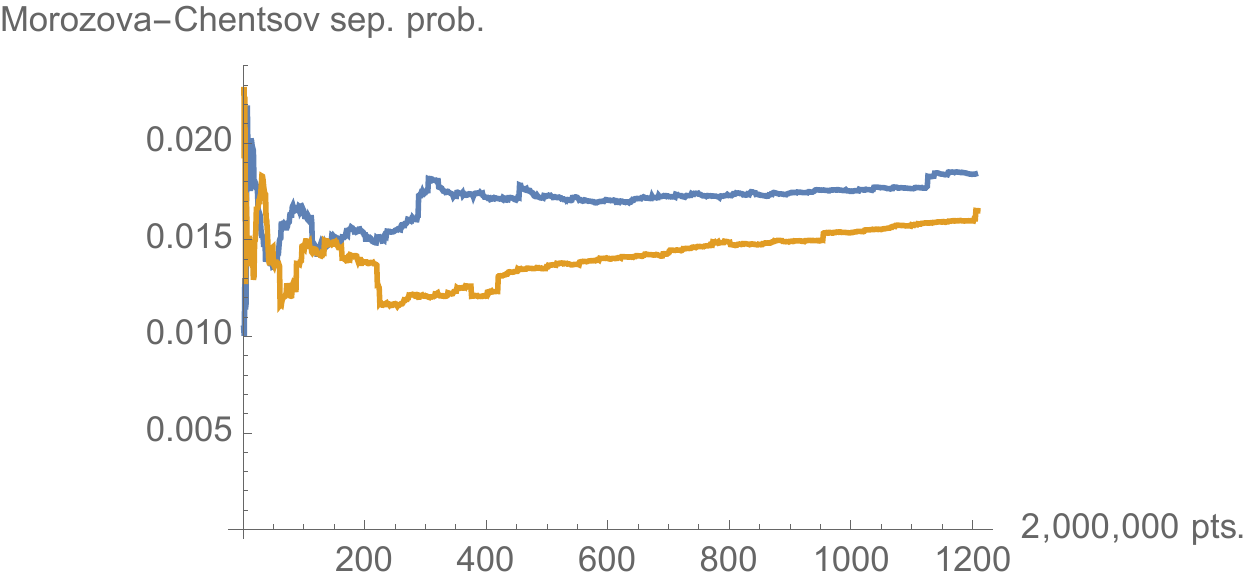}
    \caption{Pair of estimates with respect to the  Morozova-Chentsov ($f(x)=\frac{2 (x-1)^2}{(x+1) \log ^2(x)}$) measure \cite{tonchev2016monotone}}
    \label{fig:other2}
\end{figure}

In Fig.~\ref{fig:other2} we show the pair of estimates with respect to the Morozova-Chentsov ($f(x)=\frac{2 (x-1)^2}{(x+1) \log ^2(x)}$)  measure \cite[sec. II.B]{tonchev2016monotone}.

\begin{figure}
    \centering
    \includegraphics{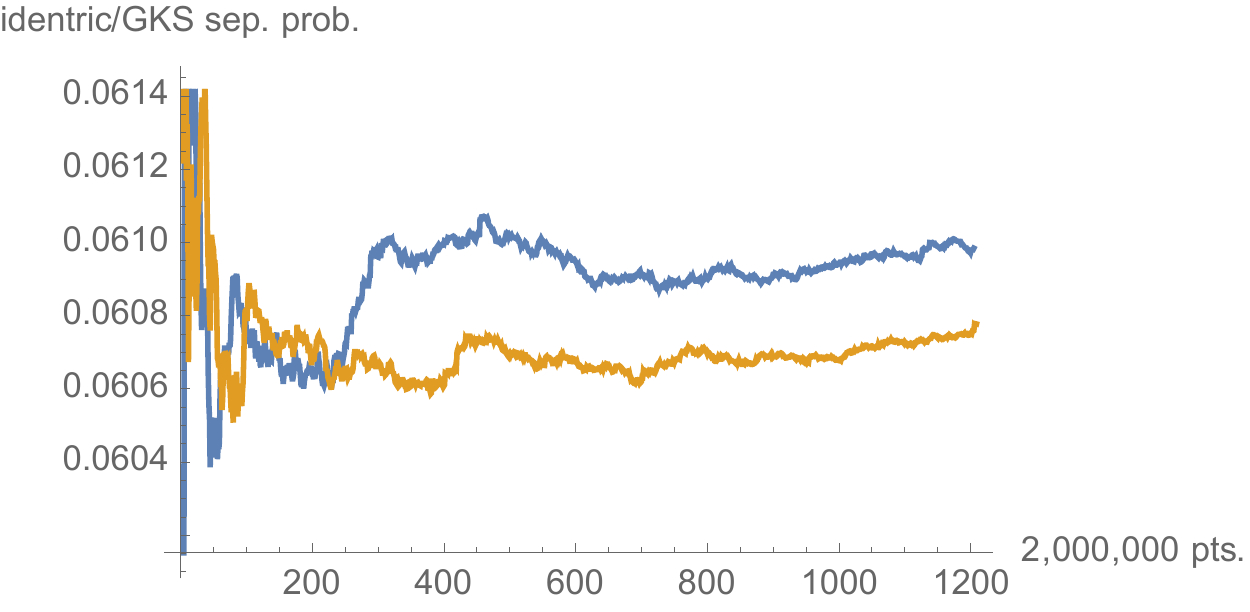}
    \caption{Pair of estimates with respect to the Grosse-Krattenthaler-Slater  (GKS)/quasi-Bures  ($f(x)=\frac{x^{\frac{x}{x-1}}}{e}$) measure}
    \label{fig:GKS}
\end{figure}

Then, in Fig.~\ref{fig:GKS} we display the pair of estimates with respect to the ``Grosse-Krattenthaler-Slater''  (GKS/quasi-Bures)  ($f(x)=\frac{x^{\frac{x}{x-1}}}{e}$)  
measure--also more broadly termed the ``identric" measure. ($\frac{2}{33} \approx 0.0606061$ is a closely-fitting value to the estimates). This mean appears to play an important role in universal quantum coding \cite[sec. IV.B]{krattenthaler2000asymptotic} \cite{slaterPRA}, in yielding the common asymptotic minimax and maximin redundancy. 

So, at this point in time, we have strongly compelling--yet no formal proof--that the Hilbert-Schmidt two-qubit separability probability is $\frac{8}{33}$ \cite{slater2018master,slaterJModPhys}, and interesting numerical evidence pointing to the Bures counterpart being $\frac{25}{341}$ \cite{slater2019numerical}.  Further, the Wigner-Yanase probability appears to be quite close to $\frac{1}{20} =0.05$. Also, we indicate that $\frac{1}{21}$ and $\frac{2}{33}$ provide close-fitting values in the arithmetic and identric cases.

If we standardize our estimate of the Bures total (separable plus entangled) volume of two-qubit states to equal 1, then the accompanying estimate of the Kubo-Mori volume is 60.7832 as large, of the Wigner-Yanase volume 7.69711 as large, and the identric/GKS volume, 2.87957 as large.  In the single-qubit case, Andai gives the Bures, Kubo-Mori, Wigner-Yanase and Morozova-Chentsov volumes as $\pi^2$, $2 \pi^2$, $4 \pi (\pi-2)$ and $\frac{\pi^4}{2}$, respectively. Based on the list of single-qubit volumes following Corollary 1 in \cite{andai2006volume}, we would anticipate that the maximal, 
geometric and 
$\frac{2 (x-1) \sqrt{x}}{(x+1) \log (x)}$-based volumes are all infinite. Along such lines, the estimates of how much larger they are than the Bures that we obtained were  $5.38871 \times 10^{18}$,  $2.80034 \times 10^{7}$ and $4.65758 \times 10^{10}$, respectively.

Upon re-examination of the detailed argument of Lovas and Andai \cite{lovasandai}, in particular their Corollary 3, we considered the possibility that rather than the geometric-mean ($\sqrt{x}$-based) two-qubit conjecture (\ref{sepX1}), we might have (again with the random-matrix 
Dyson-index $d$ set to 2) the formula (replacing the four occurrences in (\ref{sepX1}) of $-d/4$ with $\frac{d-2}{4}$)
\begin{equation} \label{sepX2}  
\mathcal{P}_{sep.\sqrt{x}}(\mathbb{C})=\frac{\int\limits_{-1}^1\int\limits_{-1}^x  \tilde{\eta}_{d} \left(
\left.\sqrt{\frac{1-x}{1+x}}\right/ \sqrt{\frac{1-y}{1+y}}	
	\right)\left(1-x^2\right)^{\frac{d-2}{4}} \left(1-y^2\right)^{\frac{d-2}{4}} (x-y)^d \mbox{d} y\mbox{d} x}{\int\limits_{-1}^1\int\limits_{-1}^x \left(1-x^2\right)^{\frac{d-2}{4}} \left(1-y^2\right)^{\frac{d-2}{4}} (x-y)^d \mbox{d} y \mbox{d} x}=
\end{equation}
\begin{equation} \label{sqrtxconjecture2}
\frac{-\frac{4}{27} \left(60 \pi ^2-593\right)}{\frac{4}{3}}= \frac{1}{9} \left(593-60 \pi ^2\right) \approx 0.0915262.
\end{equation}
(We note that 593 is prime.) For the two-rebit [$d=1$] case, the two formulas are simply equivalent--that is, $-d/4=\frac{d-2}{4}=-\frac{1}{4}$.  Also, both these conjectures assume that the formally proven result $\tilde{\chi}_{1}(\varepsilon) = \tilde{\eta}_{1}(\varepsilon)$ \cite[Lemma 7, App. B]{lovasandai} can be extended to  the proposition that $\tilde{\chi}_{2}(\varepsilon) = \tilde{\eta}_{2}(\varepsilon)$. For $d=2$, the terms $(1-x^2)$ and $(1-y^2)$ simply ``disappear'' from the integrands in (\ref{sepX2})--an apparent further manifestation of simplification in the standard 15-dimensional convex set of two-qubits framework.

A separability probability as large as 0.0915262 did seem somewhat somewhat surprising to us, as we had come to believe that the Bures ({\it minimal} monotone) two-qubit one--conjectured to be $\frac{25}{341} \approx 0.0733138$--is the largest among the family of operator monotone measures. Continuing with this  $\frac{d-2}{4}$-{\it ansatz}, the two-quaterbit separability probability--using (\ref{quater})--would then be the ratio of $\frac{3342341 \pi ^2}{64}-\frac{1136525312}{2205}$ to $\frac{5 \pi ^2}{64}$, that is, 
$\frac{3342341}{5}-\frac{72737619968}{11025 \pi ^2} \approx 0.014015$. (We have $72737619968=2^{23} \cdot 13 \cdot 23 \cdot 29$ and $11025 =3^2 \cdot 5^2 \cdot 7^2=105^2$, while 3342341 is itself prime.)

It would certainly be a lofty goal to seek a higher-order function (``functional") $f$ that given any operator monotone function would return the corresponding two-qubit separability probability. In regard to such a line of thought, J. E. Pascoe wrote: ``It might be useful to consider the fact that operator monotone functions are exactly self maps of the upper half plane, and therefore have nice integral representations. In Peter Lax `Functional Analysis' book, I think these are called `Nevanlinna representations'.  To make a long story short, this would make your function $f$ depend on a real number $a$, a nonnegative $b$ and a positive measure on the real line $\mu$.''

\appendix 
\section{Absolute separability probabilities} \label{appendix}
In \cite[eq. (34)]{JMP2008}, making use of the eigenvalue inequality formula 
\cite[eq. (3)]{hildebrand2007positive},
\begin{equation}
\lambda_1 \leq \lambda_3 +2 \sqrt{ \lambda_2 \lambda_4}    
\end{equation}
we reported a formula for the Hilbert-Schmidt two-qubit {\it absolute} separability probability \cite{kus2001geometry,arunachalam2014absolute}--measuring the proportion of states that can not be entangled by unitary transformations--of the 15-dimensional convex set of two-qubit states. It was later further condensed to 
\begin{equation} \label{HSabs}
 \frac{29902415923}{497664}+\frac{-3217542976+5120883075 \pi -16386825840 \tan
   ^{-1}\left(\sqrt{2}\right)}{32768 \sqrt{2}} =
\end{equation}
\begin{displaymath}
\frac{32(29902415923 - 24433216974 \sqrt{2})+248874917445 \sqrt{2}(5  \pi - 16  \tan ^{-1}\left(\sqrt{2}\right))}{2^{16} \cdot 3^5} \approx 0.00365826,
\end{displaymath}
much smaller than the combined (absolute and non-absolute) separability probability of $\frac{8}{33} \approx 0.242424$.
(``[C]opious use was made of trigonometric identities involving the tetrahedral dihedral angle $\phi=\cos ^{-1}\left(\frac{1}{3}\right)$'', assisted by V. Jovovic. Equation (\ref{HSabs}) here corrects a misprint in eq.  (A2) in \cite{slater2018master}. We also confirmed this highly challenging-to-obtain 2009 result,  at least to high numerical precision, in a {\it de novo} analysis.)

In \cite[sec. III.C]{JMP2008}, we also gave a Bures two-qubit absolute separability probability estimate of 0.000161792. 
(Startingly, in essentially total agreement with these last two results, in \cite[Table 2]{khvedelidze2015geometric}, Khvedelidze and Rogojin reported Hilbert-Schmidt and Bures estimates of 0.00365826 and 0.000161792, respectively.) 

\begin{figure}
    \centering
    \includegraphics{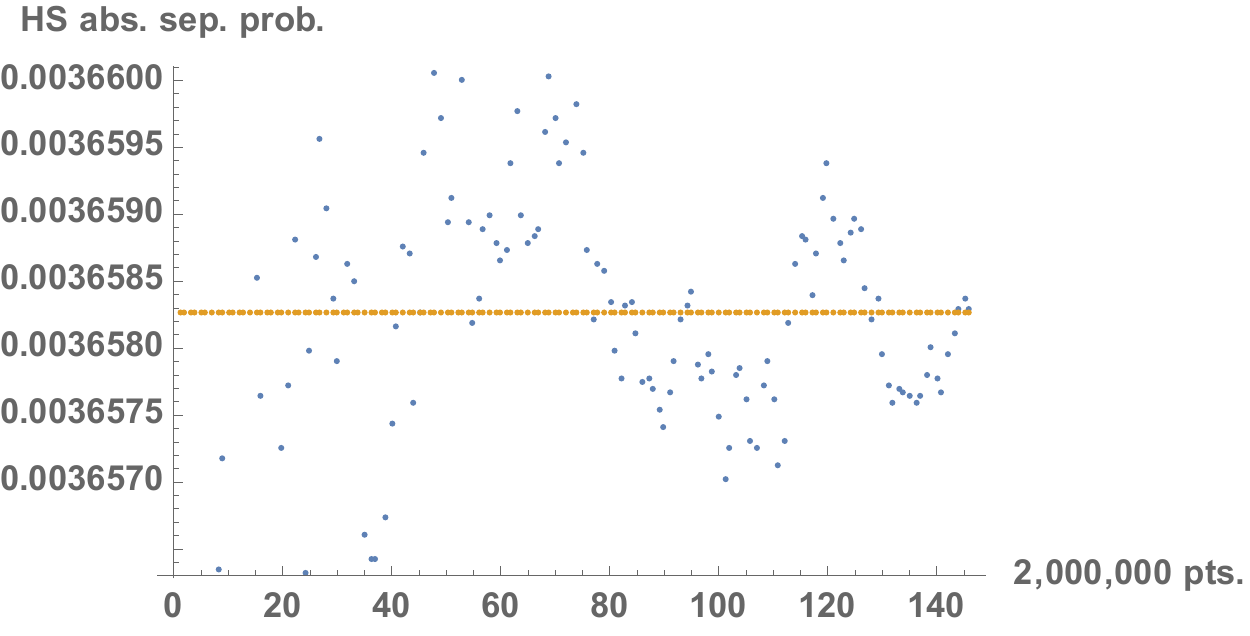}
    \caption{Quasirandom estimation (with the Roberts parameter set to $\alpha_0=0$) of the Hilbert-Schmidt two-qubit absolute separability probability along with the predicted value (\ref{HSabs}).}
    \label{fig:HSAbsolute}
\end{figure}

In certain of our 15-dimensional quasirandom estimations conducted earlier here, we also collaterally estimated  the lower (4)-dimensional absolute separability probabilities (rather than in a {\it de novo} 4-D analysis). For instance, 
in Fig.~\ref{fig:HSAbsolute}, we now show our quasirandom estimation (with $\alpha_0=0$) of the Hilbert-Schmidt two-qubit absolute separability probability along with the predicted value (\ref{HSabs}).

\begin{figure}
    \centering
    \includegraphics{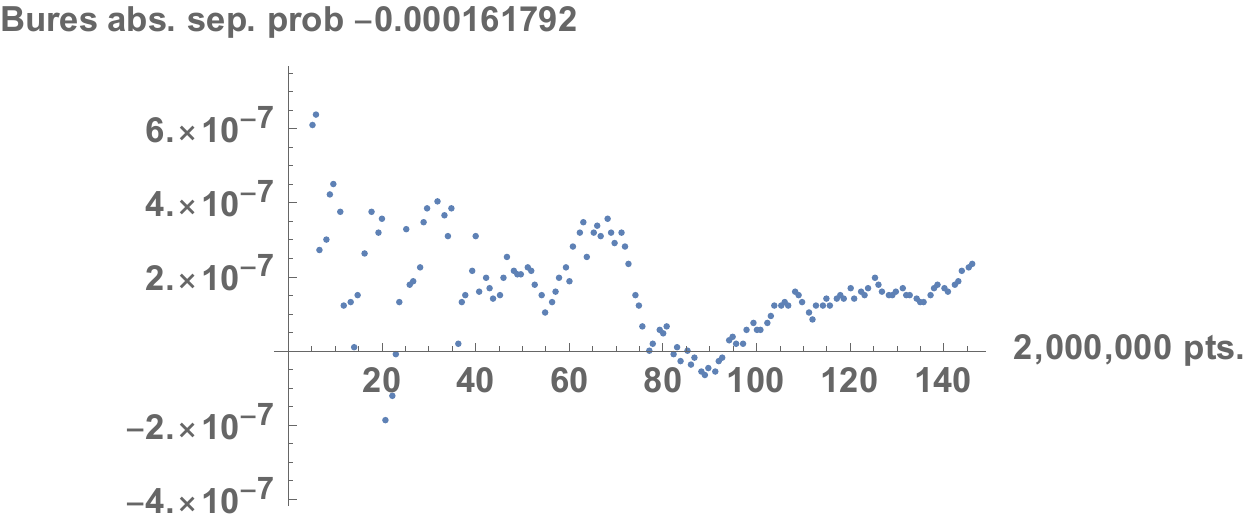}
    \caption{Deviations about the previously tabulated value of 0.000161792 of a quasirandom estimation (with $\alpha_0=0$) of the Bures two-qubit absolute separability probability}
    \label{fig:BuresAbsolute}
\end{figure}

In Fig.~\ref{fig:BuresAbsolute}, we show the deviations about the--as 
indicated--previously tabulated value of 0.000161792 of a quasirandom estimation (with $\alpha_0=0$) of the Bures two-qubit absolute separability probability.

\begin{figure}
    \centering
    \includegraphics{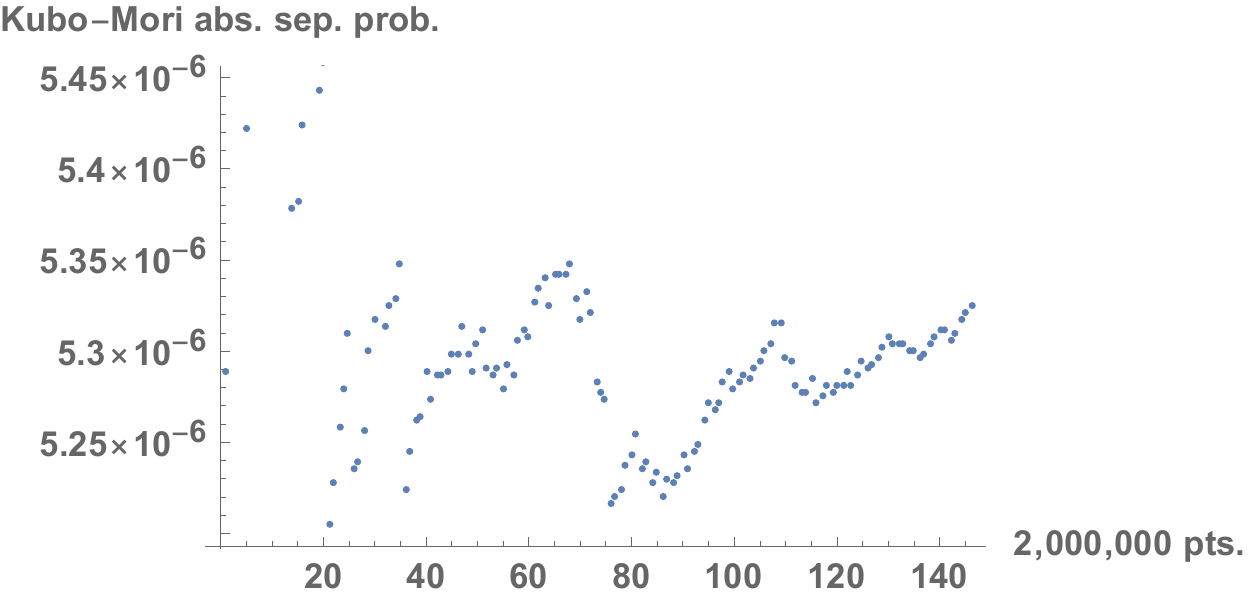}
    \caption{Quasirandom estimation (with $\alpha_0=0$) of the Kubo-Mori two-qubit absolute separability probability}
    \label{fig:KuboMoriAbsolute}
\end{figure}

\begin{figure}
    \centering
    \includegraphics{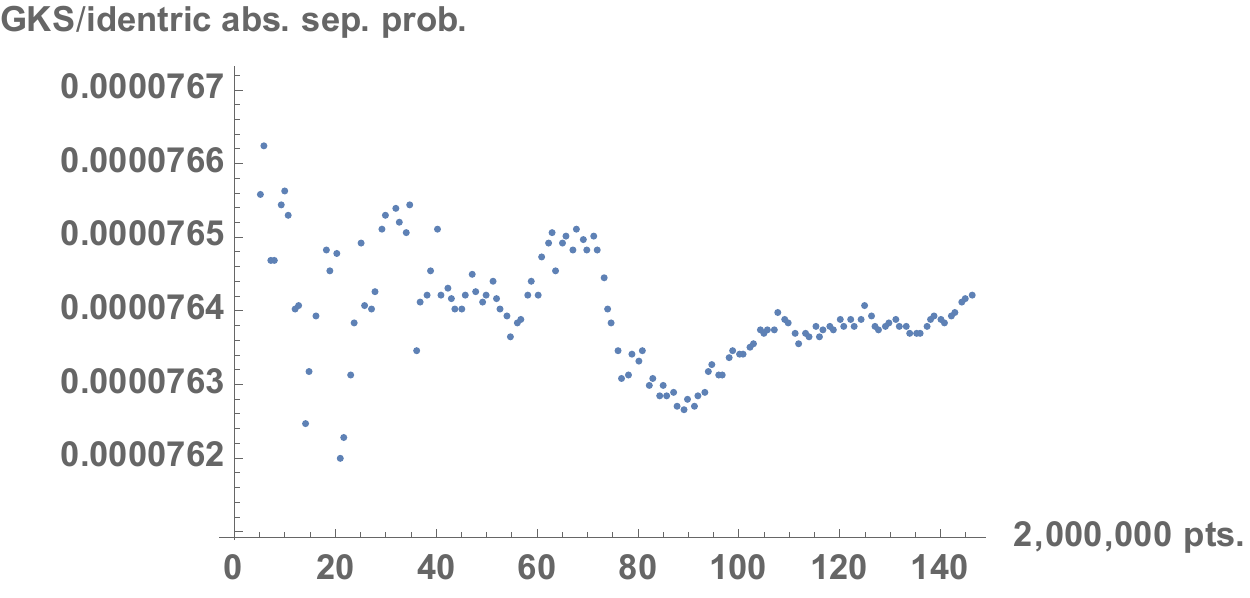}
    \caption{Quasirandom estimation (with $\alpha_0=0$) of the GKS/identric two-qubit absolute separability probability}
    \label{fig:GKSAbsolute}
\end{figure}

In Fig.~\ref{fig:KuboMoriAbsolute}, we show a quasirandom estimation (again with $\alpha_0=0$) of the Kubo-Mori two-qubit absolute separability probability
and in Fig.~\ref{fig:GKSAbsolute}, we present a quasirandom estimation (with $\alpha_0=0$) of the GKS/identric two-qubit absolute separability probability.

Our last (presumably most precise) quasirandom estimates of the absolute separability probabilities with respect to the Kubo-Mori, Wigner-Yanase and identric measures are $5.31648 \times 10^{-6}$, 0.0000343464 and 0.000076423, respectively.

Independent 4-dimensional, more conventional-type, numerical integrations gave $5.04898 \times 10^{-6}$, $0.0000342309$  and 0.0000762634 for the Kubo-Mori, Wigner-Yanase and identric absolute separability probabilities.

For the $k=1$ case of induced measure ($k=0$ corresponding to the Hilbert-Schmidt instance), for which the two-qubit separabilty probability is $\frac{61}{143} =\frac{61}{11 \cdot 13} \approx 0.426573$ \cite{slater2018qubit}, the absolute separability probability is $\approx 0.0232545$. For $k=2$, the corresponding pair of probabilities is $\frac{259}{442} $ and $\approx 0.071066971$. For $k=3,4$, the absolute separability probabilities increase substantially to approximately
0.1499309 and 0.252828. 

\begin{figure}
    \centering
    \includegraphics{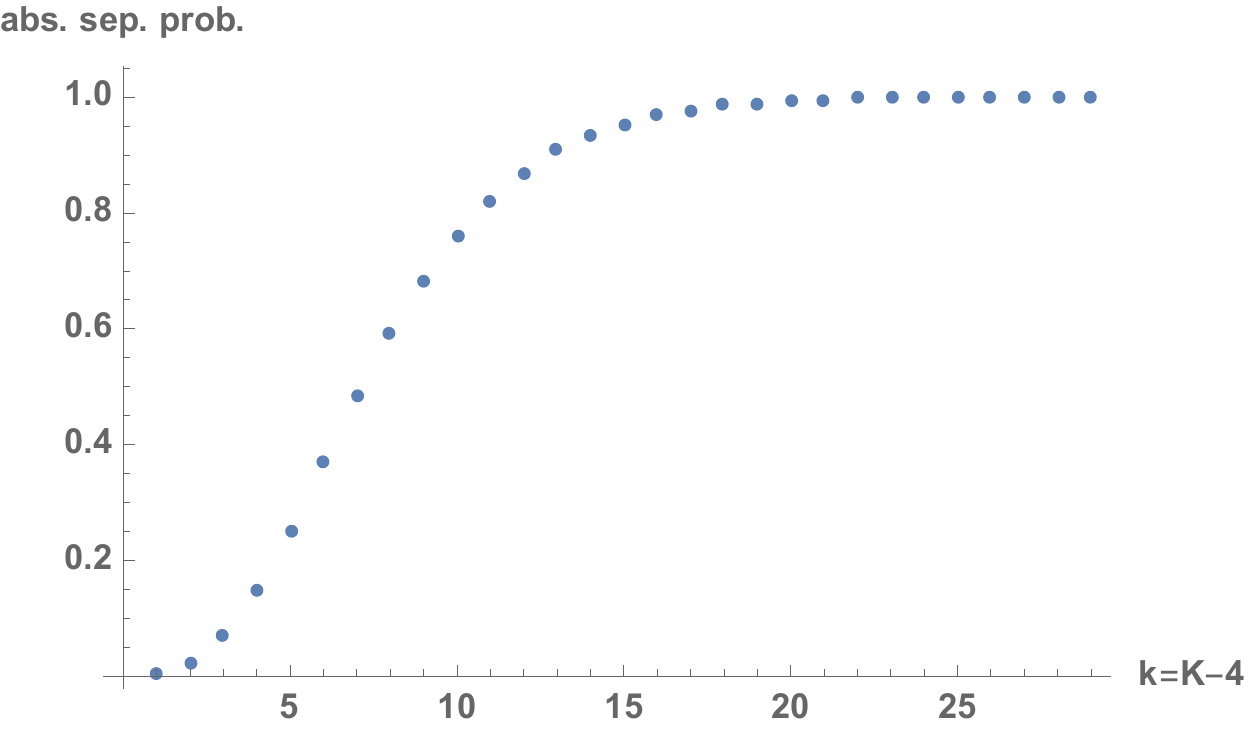}
    \caption{Increase in the absolute separability probability as the induced measure parameter $k=K-4$ increases from the Hilbert-Schmidt value of $k=0$, at which the probability is given by (\ref{HSabs})}
    \label{fig:InducedAbsSepProb}
\end{figure}

In Fig.~\ref{fig:InducedAbsSepProb} we plot the  absolute separability probability as the induced measure parameter $k=K-4$ ($N=4$) increases from the Hilbert-Schmidt setting of $k=0$, at which the probability is given by (\ref{HSabs}).
(``The natural, rotationally invariant measure on the set of all pure states of a $N \times K$  composite system, induces a unique measure in the space of $N \times N$  mixed states'' \cite{Induced}. The parameter $k$ is the difference [$k= K-N$] between the dimensions [$K,N$,with $K\geq N$] of the subsystems of the pure state bipartite system in which the density matrix is regarded as being embedded \cite{Induced}.)
\subsection{Variation with Bloch radius of qubit subsystems}
\begin{figure}
    \centering
    \includegraphics{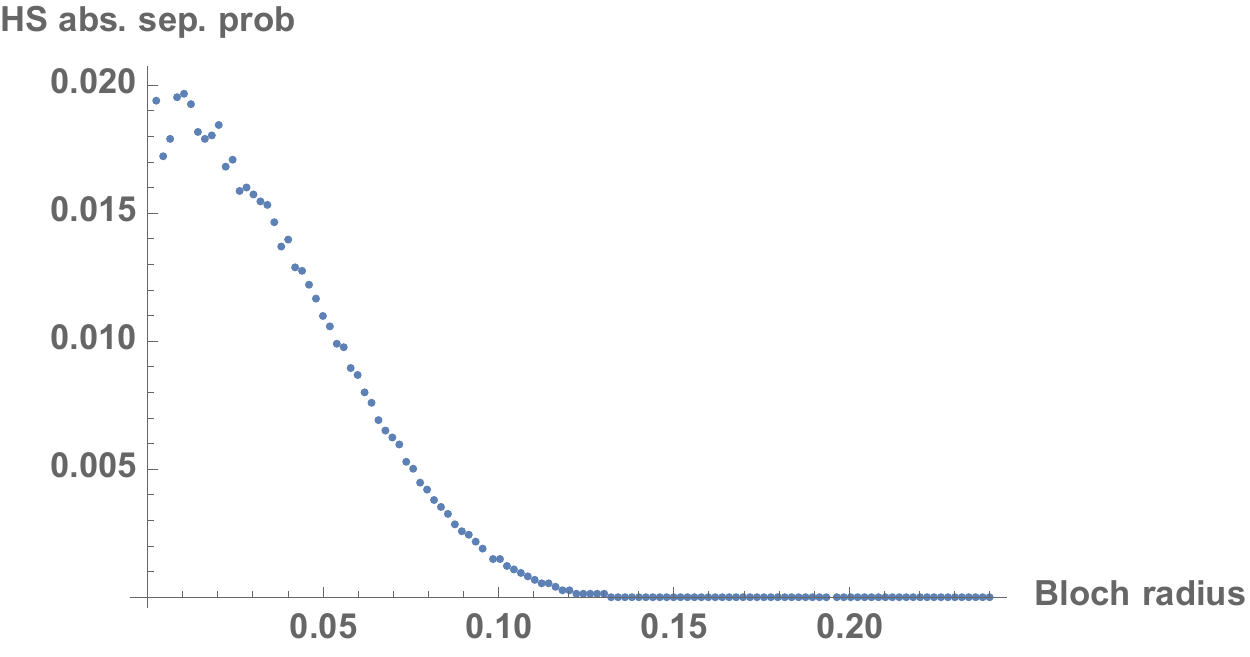}
    \caption{Hilbert-Schmidt two-qubit {\it absolute} separability probability as a function of the Bloch radii of the reduced qubit subsystems. Without the absoluteness requirement, the curve is flat at $\frac{8}{33}$.}
    \label{fig:BlochRadiusAbsSepProb}
\end{figure}

In Fig.~\ref{fig:BlochRadiusAbsSepProb} we show the Hilbert-Schmidt two-qubit {\it absolute} separability probability--given by (\ref{HSabs})--as a function of the Bloch radii of the reduced qubit subsystems. In the (total/absolute and non-absolute) Hilbert-Schmidt separability probability case--by results of Lovas-Andai and Milz-Strunz \cite{lovasandai,milzstrunz}--the corresponding curve is {\it flat} at the value of $\frac{8}{33}$. (An effort to produce a corresponding plot in the qubit-qutrit case--where the eigenvalue condition $\lambda_1-\lambda_5 -2 \sqrt{\lambda_1 \lambda_6} \leq 0$ would be implemented--proved somewhat problematical as realizations, meeting this 
requirement--of absolutely separable states were very rare.)

\begin{acknowledgements}
This research was supported by the National Science Foundation under Grant No. NSF PHY-1748958.
\end{acknowledgements}

\bibliography{main}% Produces the bibliography via BibTeX.

\end{document}